\newlength{\extralineskip}
\documentclass[12pt]{article}

\usepackage{amssymb}

\newcommand{\be}{\begin{equation}}
\newcommand{\ee}{\end{equation}}

\newcommand{\ba}{\begin{eqnarray}}
\newcommand{\ea}{\end{eqnarray}}

\newcommand{\keV}{{\rm keV} }

\begin{document}
\begin{titlepage}
\begin{flushright}
          \begin{minipage}[t]{12em}
          \large UAB--FT--556\\
                 November 2003
          \end{minipage}
\end{flushright}
\vspace{\fill}

\vspace{\fill}

\begin{center}
\baselineskip=2.5em

{\large \bf Axions}
\end{center}

\vspace{\fill}

\begin{center}
{\bf  Eduard Mass\'o}\\
\vspace{0.4cm}
     {\em Grup de F\'\i sica Te\`orica and IFAE\\
     Universitat Aut\`onoma de Barcelona\\
     08193 Bellaterra, Barcelona, Spain}
\end{center}
\vspace{\fill}

\begin{center}
\large Abstract
\end{center}
\begin{center}
\begin{minipage}[t]{36em}
I review the physics of axions, paying attention to
the role as dark matter.
This paper is based on talks given at the workshops
``Thinking, Observing and Mining the Universe''
held in Sorrento (Italy), September 22-27, 2003,
and at
``International Workshop on Astroparticle and High Energy Physics''
held in Valencia (Spain) October 14-18, 2003.
\end{minipage}
\end{center}

\vspace{\fill}

\end{titlepage}

\clearpage

\addtolength{\baselineskip}{\extralineskip}

\section{Introduction}

The QCD Lagrangian contains the so-called $\theta$-term
\begin{equation}
 \theta_{QCD} \, \frac{\alpha_s}{8\pi}\
G^a_{\mu\nu} \widetilde G^{a\mu\nu}\ + ...
\end{equation}
with
$\widetilde G^a_{\mu\nu}= (1/2) \epsilon_{\mu\nu\rho\sigma}
G^{a\rho\sigma}$, which
is CP-violating. One may modify the
value of $\theta$ by performing axial $U(1)_A$ rotations
on the quarks. In turn, these rotations change the
value of the phase of the determinant of the quark mass matrix
$M$. In fact, there is a combination that is invariant under
$U(1)_A$ rotations,
\begin{equation}
  \overline \theta = \theta_{QCD} + {\rm Arg\ Det}\ M
\end{equation}

A non-zero $\overline \theta$ would imply CP violation, most notably it
would endow the neutron with an electric dipole moment
\begin{equation}
d_n  \sim \frac{e}{m_n}\   \overline \theta   \
\frac{m_u m_d}{m_u + m_d}\,
\frac{1}{\Lambda_{QCD}}
\end{equation}
The experimental bound on this observable,
\begin{equation}
d_n < 0.63 \times 10^{-25}\ e\, {\rm cm}
\end{equation}
leads to
\begin{equation}
  \overline \theta < 10^{-9}
\end{equation}
Why is  $\overline \theta$  so small?
We would have expected   $\theta_{QCD}$ and Arg Det $M$
not far from  O(1), and we have no reason to expect
such fine-tuned cancellation between  the two terms
$\theta_{QCD}$ and Arg Det $M$, since they have totally unrelated
origins. This is the so-called strong CP-problem.

Peccei and Quinn found a solution \cite{PecceiEM}
to the strong CP problem by
introducing a new global chiral symmetry   $U(1)_{PQ}$
and using the freedom to rotate $\overline \theta$ away.
The spontaneous symmetry breaking of $U(1)_{PQ}$ at energy $\sim f_a$
generates a Goldstone boson: the axion, $a \sim   f_a \overline \theta$
 \cite{WeinbergEM}. We should keep in mind however that
the PQ solution to the strong CP-problem is not the unique
solution (see \cite{ChengEM} for a review).

The axion couples derivatively to matter
\begin{equation}
{\cal L }_{a\Psi\Psi} =
 \sum_i c_i\frac{1}{2   f_a }
(\bar\Psi_i \gamma^\mu \gamma_5 \Psi_i)
(\partial_\mu a)
\end{equation}
Here $i=e,p,n,$ etc, and
$c_i=O(1)$ are model dependent parameters.
The axion has also a non-derivative coupling
to two gluons,
\begin{equation}
{\cal L }_{agg} = \frac{1}{  f_a   }\
\frac{\alpha_s}{8\pi}\ G \cdot \widetilde G\ a
\end{equation}
At low ($\Lambda_{QCD}$) energies, the $gga$ term
gives rise to the potential $V(\overline \theta)$ that makes
$\overline \theta\rightarrow 0$  and also generates a mass for the axion
\begin{equation}
  m_a   =
\frac{f_\pi m_\pi}{  f_a  } \frac{\sqrt{m_u m_d}}{m_u+m_d}=
0.6\ {\rm eV}\ \frac{10^7\, {\rm GeV}}{  f_a  }
\label{famarelation}
\end{equation}
Due to these properties the axion is not exactly a Goldstone
boson (which is exactly massless and has only derivative couplings).

The axion has also an effective coupling to two photons:
\begin{equation}
{\cal L }_{a\gamma\gamma} = c_\gamma\, \frac{\alpha}{\pi f_a }\
F \cdot \tilde F\ a
= -   g_{{a\gamma\gamma}}\,   \vec E \vec B\, a
\label{axionphotonphoton}
\end{equation}
Such a coupling is important from the point of view
of a possible detection.

Let us stress that all $c_i$ are mildly model dependent
except for the electron $c_e$ parameter. Indeed, there are
models with $c_e=0$, i.e., the axion is not coupled
to $e$ at tree level (KSVZ type or ``hadronic axion'') \cite{KimEM}.
However, most models have $c_e\neq0$, for example in
GUT-embedded models like the DFSZ type \cite{DineEM}.
Let us check, as a way of example, that
the $a\gamma\gamma$ coupling is not wildly model dependent.
For the DFSZ-type axion we have $c_\gamma=0.36$ and
for the KSVZ-type we have $c_\gamma=-0.97$.

\section{Status of the axion}

The scale $f_a$ has to be a very high energy scale.
This is an empirical fact that we know when
using the constraints coming from
particle physics experiments and astrophysical
observations. Since
$f_a$ and $m_a$ are related (\ref{famarelation}),
this imply that the axion has to be very light. We also
see, from the axion couplings showed in the last section,
that it must be a very weakly interacting particle.

Particle physics experiments, namely, decays involving axions
or beam dump experiments lead to
\begin{equation}
  f_a   > 10^4\, {\rm GeV}
\end{equation}
or, equivalently,
\begin{equation}
  m_a   < 1\, \keV
\end{equation}
This excludes that $f_a$ could be on the order of the Fermi scale,
which was the original suggestion of Peccei and Quinn \cite{PecceiEM}.

Astrophysical limits are obtained when considering
that a ``too'' efficient energy drain due
to axion emission would be inconsistent  with observation.
Stringent limits come from horizontal branch stars in
globular clusters. The main production is from the Primakov
process $\gamma \gamma^* \rightarrow a$ where $\gamma^*$ corresponds
to the electromagnetic field induced by protons and electrons in
the star plasma. The coupling is restricted to \cite{RaffeltEM}
\begin{equation}
  g_{{a\gamma\gamma}}   < 0.6 \times 10^{-10}\, {\rm GeV} \hspace*{.3cm}
\Rightarrow \hspace*{.3cm}
  f_a   > 10^7\, {\rm GeV}
\end{equation}
In terms of axion mass, the interval
\begin{equation}
0.4\, {\rm eV}\ <   m_a   < 200\, \keV
\end{equation}
is ruled out (for $m_a >200\, \keV$ the axion is too
heavy to be produced).

The most restrictive astrophysical limits on the axion parameters
come from the analysis of neutrinos from the supernova
SN 1987A. In the supernova core, the main production is
axion bremsstrahlung in nucleon-nucleon processes,
$NN \rightarrow NN a $.
The observed duration of the $\nu$ signal at Earth detectors constrains
the coupling of the axion to nucleons. The range
\begin{equation}
3 \times 10^{-10}<
  g_{ann}   \equiv  c_n \frac{m_n}{  f_a   }
< 3 \times 10^{-7}
\label{limitSN}
\end{equation}
is excluded \cite{EllisEM}.
The upper limit in (\ref{limitSN}) corresponds to axion trapping
in the SN. The lower limit in (\ref{limitSN}) is equivalent to
$ f_a  > 6 \times 10^{8}$ GeV.
In terms of the axion mass, the excluded range
corresponding to (\ref{limitSN}) is
\begin{equation}
0.01\, {\rm eV}\ <   m_a   < 10\, {\rm eV}
\end{equation}

Putting together all the information coming from laboratory and
astrophysics  we may conclude that the scale
of the PQ breaking is bounded by
\begin{equation}
f_a  > 6 \times 10^{8}\ {\rm GeV}
\label{limitfa}
\end{equation}

\section{Dark matter axions}

A very exciting possibility is that, if the axion exists
and its scale is not far from (\ref{limitfa}), it may
be a substantial fraction of the dark matter of the universe.
Let us review the cosmological production of axions in the
early universe.

The cosmological history of the axion starts at temperatures
$T \sim f_a$, where $U(1)_{PQ}$ is broken. All vacuum expectation
values $<a>$ are equally likely, but naturally we expect $<a>$
of the order of the PQ scale, or in other words an
initial angle:   $\overline \theta_1 \sim <a>/f_a \sim 1  $.
At $T \sim \Lambda$ (We define $\Lambda=\Lambda_{QCD}$),
QCD effects turn on and create a potential $V(\overline \theta)$
that forces $\overline \theta \rightarrow 0$
(CP-conserving value). Thus,
the $\theta$ angle starts at $  \overline \theta=\overline \theta_1\sim 1  $
and will relax to   $\overline \theta \rightarrow 0$; we say it
was ``misaligned''.  In the process of relaxation,
the field oscillations contribute to the cosmic energy density
 \cite{PreskillEM} in the form of a population of non-relativistic
axions.

Although the larger the scale $f_a$ is, the weaker the axion interacts,
and more early in the universe the PQ phase transition occurs,
the density $\rho_a$ in axions increases with $f_a$.

The vacuum expectation value of the axion field is subject to the
evolution equation
\begin{equation}
\frac{d^2 <a>}{dt^2} + 3 H(T) \frac{d <a>}{dt} + m_a^2(t)  <a> = 0
\end{equation}
with $H=\dot R/R$ the Hubble parameter.
In the interval $ f_a > T >> \Lambda$, the Hubble
parameter $H$ is much greater than $m_a$, and $<a>$ stays
constant. The axion mass is
suppressed at high energies and switches on at the QCD scale.
Here, at a temperature $T_i \sim \Lambda$, we have
\begin{equation}
m_{ai} \simeq H(T_i) \simeq \frac{\Lambda^2}{M_P}
\label{massinitial}
\end{equation}
($M_P=M_{\rm Planck})$. After this moment,
the axion field starts oscillating around $<a>=0$.

The mass of the axion increases adiabatically, and thus the oscillation
is sinusoidal with a time-decreasing amplitude
\begin{equation}
<a>\, \simeq A(t) \cos m_a(t) t
\end{equation}
that corresponds to an axion number density
\begin{equation}
n_a \sim m_a A^2 \sim T^3
\label{numberdensity}
\end{equation}
and to an energy density $\rho_a = m_a n_a$.
We can give an order of magnitude estimate of
the energy density of axions today (subscript 0 for today)
\begin{equation}
\rho_0 \sim (m_a^2 A^2)_0 \sim m_{a0}\, (m_a A^2)_i \,
\left(\frac{T_0}{\Lambda}\right)^3
\end{equation}
where we used (\ref{numberdensity}) and  $m_{a0}$
has to be interpreted as the zero-temperature axion mass,
given in (\ref{famarelation}). Since we want an order of
magnitude estimate, we use the approximate relation
$m_a\simeq\Lambda^2/f_a$. We also use $m_{ai}$
from (\ref{massinitial}) and
put $A_i \sim <a>_i \sim f_a$. Finally,
\begin{equation}
\rho_0 \sim \frac{\Lambda^2}{f_a}
\frac{\Lambda^2}{M_P} f_a^2
\left(\frac{T_0}{\Lambda}\right)^3
\sim f_a \frac{T_0^3 \Lambda}{M_P}
\end{equation}
and indeed we verify that the relic density is an increasing
function of $f_a$.

A careful estimate of the this energy density gives \cite{TurnerEM}
\begin{equation}
  \Omega h^2
\simeq 2 \times 10^{\pm 0.4}\, F(\overline \theta_1) \overline \theta_1^2 \
\left( \frac{10^{-6}\, {\rm eV}}{  m_a   } \right)^{1.18}
\end{equation}
where as usual, $\Omega$ is the density normalized to the
critical density, and $h$ is the Hubble parameter today in
units of 100 km/s/Mpc.
$F$ takes into account an-harmonic effects.

We see that a
 potentially interesting range for cosmology is
\begin{equation}
  \Omega h^2 \sim 1-0.1  \ \ \Rightarrow \ \
   m_a  \sim 10^{-3} - 10^{-6}\, {\rm eV}
\end{equation}
since then the axion could be part of the cold dark matter
of the universe.

If we have as initial condition $F(\overline \theta_1) \overline \theta_1^2 \sim 1$, we get a
lower bound on the axion mass
\begin{equation}
10^{-6}\, {\rm eV}\, <\,    m_a
\end{equation}
However, for smaller values of the initial $\overline \theta_1$, one gets a
looser bound. Apart from the value of $\overline \theta_1$, there are
other cosmological uncertainties.

Another axion source is the string-produced axions.
Unless inflation occurs at $T< f_a$,
axion strings survive and decay into axions. For many years,
there has been
a debate on the importance of the string mechanism, and the question
is not yet settled. On the one hand, some authors find that
$\Omega_{\rm string} \sim \Omega_{\rm misalign}$ but some other
authors find
$\Omega_{\rm string} \sim 10 \, \Omega_{\rm misalign}$.
Also, axion domain walls constitute another potential
axion source. For a review and discussion on
axionic strings and walls, see ref.\cite{SikiviereviewEM}.

\section{Axion experiments}

The effort to detect axions is based in the following
phenomena \cite{SikivieEM}. The interaction
term (\ref{axionphotonphoton}) leads to
axion-photon mixing  in an external magnetic field. In addition,
the probability of the $a -\gamma$ transition is
enhanced when the $a -\gamma$ conversion in the magnetic
field is coherent.

For example,
in the presence of a flux of galactic halo axions, we expect
conversion of axions into photons
in a cavity with a strong magnetic field (haloscope)
\cite{SikivieEM}.
When the (tunable) frequency of a cavity mode equals
the axion mass,
\begin{equation}
h\nu=E \simeq m_a (1 + \beta^2 /2) \hspace*{1cm} \beta \sim 10^{-3}
\end{equation}
there is a resonant conversion into $\mu$-wave
photons (1 GHz = 4 $\mu$eV).
Axions are supposed to be virialized in the halo with
$\beta \sim 10^{-3}$, so that there should be a very small dispersion.

Earlier experiments \cite{WuenschEM} put some limits, and presently
there is already a second-generation experiment running,
the US large scale experiment \cite{AsztalosEM},
sensitive in the range
\begin{equation}
2.9\ \mu{\rm eV}\ <   m_a   <\ 3.3\ \mu{\rm eV}
\end{equation}
This experiment has already excluded the possibility that
KSVZ axions
constitute the whole of the galactic dark matter density,
\begin{equation}
\rho = 7.5 \times 10^{-25}\ {\rm g\, cm}^{-3}
\end{equation}
In the near future, they expect to reach
$1\, \mu{\rm eV}<   m_a   < 10\, \mu{\rm eV}$
\cite{AsztalosEM}.

A promising experiment in development is CARRACK \cite{YamamotoEM},
in Kyoto. They will use a Rydberg atoms' technique
to detect $\mu$-wave photons.

Another alternative to detect axions makes use of
the flux that the Sun would emit if axions exist,
again making use of a strong
magnetic field (helioscope) \cite{SikivieEM}. The produced
photons would have energies corresponding to the Sun interior,
$E \sim $ a few keV (X-rays).

In Tokyo there is an experiment currently running, that has detected
no signal, giving the limit \cite{MoriyamaEM}
\begin{equation}
  g_{{a\gamma\gamma}}   <  6 \times 10^{-10}\ {\rm GeV}^{-1}
\end{equation}
which is only valid for $m_a   < 0.03\, {\rm eV}$, to preserve coherence.
The Tokyo experiment has been improved recently by
using gas to generate a plasmon mass
$\omega_{\rm pl}$ and thus enhancing a possible signal for
higher particle masses. They get \cite{InoueEM}
\begin{equation}
  g_{{a\gamma\gamma}}   <  6 \times 10^{-10}\ {\rm GeV}^{-1}
\end{equation}
now valid for $0.05\, {\rm eV} <  m_a < 0.26 \, {\rm eV}$.

The CAST experiment at CERN
\cite{CASTEM} is currently searching for axion from
the Sun, and is supposed to lead either to axion
discovery or to very stringent limits.

An alternative way to convert axions from the Sun
into photons is to use a crystal \cite{PaschosEM}.
Some experiments have lead to limits on the axion
parameters \cite{AvignoneEM}.

Finally, we mention the axion
laser searches, that make use of the effects
on laser propagation that axions would produce.
Stringent experimental limits are obtained
\cite{SemertzidisEM}.
Along the same lines, there is an interesting proposal
using the HERA tunnel
\cite{Ringwald:2003ns}.

\section{Axion-like particles}

As it is well known, a potential
\begin{equation}
V=V_{\rm sym}
\end{equation}
having an exact global symmetry, when
spontaneously broken implies the appearance of (massless) Goldstone
bosons. Example are
family symmetries and lepton-number symmetry,
that would  give rise to familons and majorons.

However, global symmetries are expected to be broken
by gravitation, since
black holes do not conserve global charges. The total potential
would read in this case
\begin{equation}
V=V_{\rm sym} + V_{\rm non-sym}
\end{equation}
with  $V_{\rm non-sym}$ suppressed by $M_P$ inverse powers.

In recent work we have been studying about the consequences
of the spontaneous breaking of the symmetry in presence
of (small) gravitationally-induced explicit breaking,
particularly in phase transitions in the early universe.

As a simple case, take a global $U(1)$, with a scalar field $\Psi$
\begin{equation}
V_{\rm sym}= \lambda\, [|\Psi|^2 - v^2]^2
\end{equation}
and add a small breaking
\begin{equation}
V_{\rm non-sym}= \frac{g}{M_P^{n-2}}\, |\Psi|^n \Psi^2 + h.c.
\end{equation}
with $n \geq 3$.

It is still convenient to parameterize
\begin{equation}
\Psi = [v+\rho]\exp{(i\theta/v)}
\end{equation}
We find \cite{MassoEM} that
$\rho$ decays quickly at the phase transition, but
$\theta$ starts oscillating later and gives rise
to non-relativistic massive particles.
For some range of parameters these pseudo-Goldstone
bosons may constitute dark matter.
In general, all these bosons couple
to photons, so that they could give signatures in most of the experiments
that look for axions .

\section{Conclusions}

The axion is a particle that is theoretically motivated,
since is the consequence of the Peccei-Quinn solution to
the strong CP problem. Axion physics is a very
rich subject, particularly since
it is possible that it constitutes
part of the universal dark matter.
Upcoming experiments may find the axion, or may exclude it.
Other massive pseudo-goldstone bosons have also
interesting phenomenologies.

\end{document}